\documentclass[a4paper,11pt]{article}
\pdfoutput=1 
\usepackage{jheppub}

\usepackage{graphicx}
\usepackage{dcolumn}
\usepackage{bm}
\usepackage{epstopdf}
\usepackage{mathrsfs}
\usepackage{amssymb,amsmath,amsfonts,latexsym}
\usepackage{nicefrac}
\usepackage[colorlinks=true,linktocpage=true,linkcolor=blue,citecolor=blue]{hyperref}
\usepackage[utf8]{inputenc}
\usepackage{lipsum}
\usepackage{nicefrac}
\usepackage{slashed}
\usepackage{mathtools}
\def\bs{\boldsymbol} 
\def\del{\partial}
\def\bdel{\bs\partial}
\newcommand{\eqn}[1]{Eq.~\eqref{#1}}

\long\def\comment#1{ }

\def\for{\quad\text{for}\quad} 
\def\and{\qquad\text{and}\qquad} 
\def\el{\text{el}} 
\def\br{\text{br}} 
\def\eff{\text{eff}} 
\def\BH{\text{BH}} 
 
\def\scat{\text{scatt}} 
\def\coh{\text{coh}} 
\def\mfp{\text{mfp}} 
\def\GLV{\text{GLV}} 
\def\scatt{\text{scatt}} 
\def\coh{\text{coh}}

\def\0{{\boldsymbol 0}}
\def\q{{\bm q}}

\def\k{{\boldsymbol k}}

\def\x{{\boldsymbol x}}
\def\y{{\boldsymbol y}}
\def\p{{\boldsymbol p}}

\def\r{{\boldsymbol r}}
\def\z{{\boldsymbol z}}

\def\qin{{\hat q_0}}

\def\cl{\text{cl}}

\def\Cot{\text{Cot}}
\def\Tan{\text{Tan}}

\def\HO{\text{HO}}
\def\pert{\text{pert}}

\def\Kc{{\cal K}}

\newcommand{\beq}{\begin{eqnarray}}
\newcommand{\eeq}{\end{eqnarray}}
\newcommand{\be}{\begin{eqnarray*}}
\newcommand{\ee}{\end{eqnarray*}}
\newcommand{\bal}{\begin{align}}
\newcommand{\eal}{\end{align}}
\newcommand{\rmd}{{\rm d}}

\newcommand{\rme}{{\rm e}}

\def\rmR{{\rm Re}}

\def\abar{{\rm \bar\alpha}}

\newcommand{\nn}{\nonumber\\ }

\begin{document}

\title{Gluon bremsstrahlung in finite media beyond multiple soft scattering approximation}

\author[a]{Yacine Mehtar-Tani}

\affiliation[a]{Physics Department, Brookhaven National Laboratory, Upton, NY 11973, USA}
\emailAdd{mehtartani@bnl.gov}

\date{\today}
\abstract{

We revisit the calculation of the medium-induced gluon spectrum in a finite QCD medium and develop a new approach that goes beyond multiple soft scattering approximation. We show by expanding around the harmonic oscillator that the first two orders encompass the two known analytic limits: single hard and multiple soft scattering regimes, valid at high and low frequencies, respectively.  Finally, we investigate the sensitivity of our results to the infrared and observe that for large media the spectrum is weakly dependent on the infrared medium scale. 
}

\date{\today}
\maketitle
\flushbottom

\section{Introduction}\label{sec:intro}

In the early 80's \cite{Bjorken:1982tu}, J. D. Bjorken suggested that the creation of a quark-gluon plasma in high energy hadronic collisions would cause the suppression of  high-pt jets. Two decades later \cite{Adcox:2001jp,Adler:2002tq}, this phenomenon was successfully observed at RHIC in the suppression of high-pt hadrons, and confirmed at the LHC with the quenching of fully reconstructed jets \cite{Aad:2010bu,Chatrchyan:2011sx}. 

Medium-induced radiative energy loss resulting from multiple final state interactions of high energy partons with the quark-gluon plasma is believed to be the dominant mechanism for jet-quenching \cite{Baier:1996sk,Baier:1996kr,Zakharov:1996fv,Zakharov:1997uu,Arnold:2002ja} (for recent reviews see \cite{Mehtar-Tani:2013pia,Blaizot:2015lma}). This inelastic process is triggered by multiple scattering centers that act coherently during the quantum mechanical radiation time leading to the Laudau-Pomerantchuk-Migdal (LPM) suppression of the gluon radiation spectrum  \cite{Landau:1953gr,Migdal:1956tc}. A quantitative description of jet quenching requires a full account of the various regimes of medium-induced gluon radiation, which currently is only achieved by numerical methods \cite{CaronHuot:2010bp,Ke:2018jem}. A closed analytic formula would be more practical in particular when the elementary radiative process is iterated to generate parton cascades in the context of a Monte Carlo event generators \cite{Schenke:2009gb,Zapp:2008gi}. In many phenomenological studies different approximations are adopted with limited control over the theoretical uncertainties associated with them. 

There are two known analytic limits: 
\begin{enumerate} 
\item {\it Multiple soft scattering approximation} (MSSA): This approximation, in which high density effects are resummed to all orders, is expected to be important for short mean-free-path $\ell_\mfp \ll L$. It is often referred to in the literature as the Baier-Dokshitzer-Mueller-Peigné-Schiff-Zakharov (BDMPS-Z) approximation\footnote{Note, however, that the formalism introduced by BDMPS-Z encompasses the single hard scattering limit as shown by  Wiedemann \cite{Wiedemann:2000za}.}. The BDMPS spectrum, generalized to include transverse momentum dependence and finite gluon energy \cite{Blaizot:2012fh,Apolinario:2014csa}, is the building block for medium-induced QCD cascade \cite{Jeon:2003gi,Blaizot:2013vha,Blaizot:2013hx}. 
\item  {\it Single hard scattering approximation} (SHSA): In this approach the medium is assumed to be dilute, that is, $\ell_\mfp \gg L$. It was first discussed by Gyulassy, Levai and Vitev (GLV) \cite{Gyulassy:2000er}, and Wiedemann \cite{Wiedemann:2000za} and was recently extended beyond soft gluon approximation to finite gluon frequency \cite{Ovanesyan:2011kn,Sievert:2018imd}. A similar approach, dubbed Higher-Twist (HT), involves further approximation, namely, that the gluon transverse momentum is much larger than the typical momentum transfer from the medium $\mu$ \cite{Wang:2001ifa}. 
\end{enumerate} 

MSSA breaks down at high enough gluon frequencies where at most one scattering occurs due to color transparency: hard gluons with high transverse momentum can only be produced by large momentum transfer from medium though with a small cross-section. It can be computed order by order in powers of coupling constant. Although it is a strongly suppressed process it dominates the mean energy loss. On the other hand, SHSA is expected to break down at low gluon frequencies. 

To remedy the lack of a unified formula for in-medium gluon bremsstrahlung, we propose in this work a new analytic approach inspired by the Molière scattering theory \cite{Moliere} (see also  Ref.~\cite{Iancu:2004bx} for a recent application to particle production in proton-nucleus collisions at forward rapidity), where instead of expanding around the first order in opacity, we split the interaction term into a soft and hard part. The former is resumed to all orders accounting for density effects and the latter is treated as perturbation that naturally accounts for the hard part of the radiation cross-section. We compute the first two orders and show that we recover the BDMPS and GLV-HT limits. 

The article is organized as follows. In Section \ref{sec:LPM}, we review the LPM effect in QCD and present heuristic derivations of the medium-induced gluon spectrum. In Section \ref{sec:spectrum-bdmps-glv}, we compute the medium induced spectrum in the multiple soft and single hard  scattering approximations. In Section \ref{sec:moliere}, we expand  the radiation spectrum around the harmonic oscillator and compute the first two orders. We then compare the result to the BDMPS and GLV spectra. In Section \ref{sec:generalization}, we generalize our approach to arbitrary gluon energies. Finally, we summarize and conclude in Section \ref{sec:summary}.

\section{LPM effect in QCD} \label{sec:LPM}

Before we discuss the medium-induced gluon spectrum in more details we shall first introduce the basic concepts underlying the QCD analog of the LPM effect. 

\begin{figure}
\center
\includegraphics[width=0.5\textwidth]{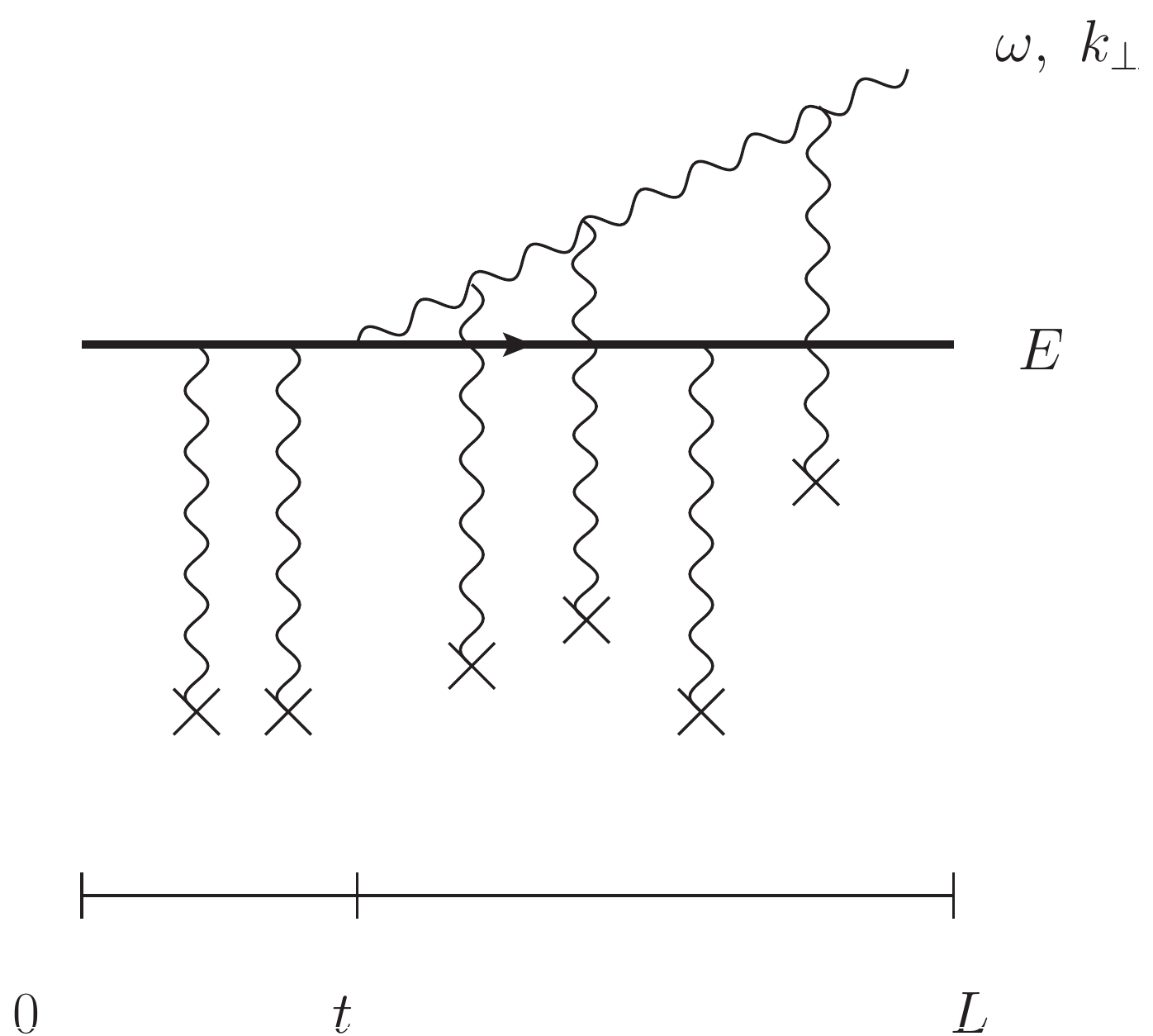}
\caption{Diagrammatic representation of a high energy quark radiating a soft gluon induced by coherent multiple scattering in a medium of length $L$. }
\label{fig:gluon-radiation}
\end{figure}

Consider a high energy parton created early in a heavy-ion collision for instance which then undergoes multiple scattering in the deconfined medium of length $L$. During its propagation, the hard parton acquires transverse momentum, $k_\perp$, w.r.t the direction of its propagation due to brownian motion in momentum space caused by random kicks from the plasma constituents. This is characterized by the transport coefficient 
\beq\label{eq:qhat-def}
\hat q \equiv \frac{\rmd \langle k_\perp^2\rangle }{\rmd t }. 
\eeq
To leading order in the strong coupling constant the quenching parameter $\hat q$ is related to the the 2 to 2 QCD matrix element  $\rmd \sigma_\el / \rmd^2 q_\perp \simeq g^4 n/q_\perp^4$, where $n$ is the density of medium color charges, as follows:
\beq\label{eq:qhat-def-lo}
 \hat q(Q^2) \simeq  C_R\,\int_\q  \, q_\perp^2 \, \frac{\rmd \sigma_\el}{  \rmd^2 q_\perp} \simeq 4\pi \alpha_s^2 C_R n \ln\frac{Q^2}{\mu^2}. 
\eeq
where $\alpha_s\equiv g^2/(4\pi)$ is the coupling constant, $C_R=C_F=(N_c^2-1)/(2N_c)$ ($C_R=C_A=N_c$) the quark (gluon) color charge and $\ln (Q^2 / \mu^2)$ the Coulomb logarithm that depends on an infrared cut-off $\mu$ which is related to the Debye screening mass in the QGP. $Q^2$ is a process dependent hard scale, that will be specified below.  Above and throughout we shall use the shorthand notation 
\beq
\int_q \equiv \frac{\rmd^2 \q}{(2\pi)^2},
\eeq
In addition to momentum broadening, multiple scatterings can trigger gluon radiation off the energetic parton as depicted in Fig.~\ref{fig:gluon-radiation}. Owing to the quantum nature of the radiation, many scattering centers act coherently during the radiation time $t_\coh = \omega/k_\perp^2$, where $\omega$ is the radiated gluon frequency and $k_\perp^2 \sim \hat q \,t_\coh$ (cf. \eqn{eq:qhat-def} ) is the transverse momentum squared acquired by the gluonic fluctuation during $t_\coh$. Solving for the coherence time and transverse momentum squared we obtain
\beq\label{eq:kt2}
t_\coh \equiv \sqrt{\frac{\omega }{\hat q }}, \and  k_{\perp,\coh}^2 \equiv \sqrt{\omega \hat q },
\eeq
respectively. In the above parametric estimate, we have assumed $\hat q$ to be constant by ignoring the slowly varying Coulomb logarithm, which must be evaluated at $Q^2 \equiv  \sqrt{\omega \hat q }$. 

Because medium-induced radiation can occur anywhere along the medium with equal probability (provided $t_\coh \ll L$ corresponding  to gluon frequency $\omega \ll \omega_c\sim \hat q L^2$), the radiation spectrum is expected to scale linearly with $L$. Therefore, dimensional analysis yields 
\beq\label{eq:bdmps-param}
\omega \frac{\rmd I}{\rmd \omega} \simeq \alpha_s\frac{L}{t_\coh} =\alpha_s \sqrt{\frac{\omega_c}{\omega}} \quad\for \quad\omega \ll \omega_c.
\eeq 
This is the BDMPS-Z spectrum in the multiple soft scattering approximation. 
Note that the above spectrum is valid so long as $\ell_\mfp \ll  t_\coh \ll L$. When $t_\coh \sim  \ell_\mfp$ only  one scattering is involved in the radiation process. This is the so-called Bethe-Heitler (BH) regime, where the spectrum scales with the number of scattering centers  $N_\scat$ 
\beq 
\omega \frac{\rmd I_\BH}{\rmd \omega} \simeq \alpha_s\frac{L}{\ell_\mfp} =\alpha_s N_\scat,
 \eeq
where $\ell_\mfp \sim (n \sigma_\el)^{-1}\sim \mu^2/(\alpha_s^2 n)$.
It corresponds to the incoherent limit of the radiation spectrum. In the BH regime the typical transverse momentum squared is of order the infrared scale, i.e.,  $k_\perp^2 \sim \hat q \ell_\mfp \sim \mu^2 $ (for a thermal medium $\mu\equiv m_D$, the Debye mass)

With increasing frequency the coherence time increases. As a result, the effective number of scattering centers decreases as $N_\eff \equiv N_\scat/ N_\coh = L/t_\coh$. This yields the suppression of the radiation spectrum that falls as $\omega^{-1/2}$ compared to the flat Bethe-Heitler spectrum shown in \eqn{eq:bdmps-param}. 

Let us turn now to the regime $\omega > \omega_c$. In this case, the transverse momentum is bounded from below by $Q_s^2\equiv \hat q L$. Moreover, the LPM effect suppresses the spectrum for formation times $\omega/k_\perp^2 \gg L$, which implies that $k_\perp^2 \gtrsim \omega/L \gg Q_s^2$, where the second inequality follows from the condition $\omega\gg\omega_c$. We see that the transverse momentum lies in the regime of single hard scattering since $k_\perp^2\gg Q_s^2$, where the cross-section falls as $k_\perp^{-4}$ and scales linearly with $g^4 n$ and $L$, namely, $Q_s^2$. Due to the steepness of the latter $k_\perp$ power spectrum, the integral over $k_\perp$ is dominated by the lower bound $\omega/L$.  Hence, we estimate the gluon spectrum integrated over $k_\perp$ as follows

\beq
\omega \frac{\rmd I}{\rmd \omega} \sim \alpha^3_s n L \int^\infty_{\omega/L}\frac{\rmd k_\perp^2}{k_\perp^4}\simeq \alpha_s\frac{\omega_c }{\omega}\quad\for\quad \omega \gg \omega_c.
\eeq
The above spectrum correspond to the UV limit of the GLV (HT) spectrum. 

In what follows, our goal will be to develop an analytic approach that accounts for these two limiting cases. 

\section{Medium-induced gluon spectrum and its two limits}\label{sec:spectrum-bdmps-glv}
Our starting point is the general expression for the medium-induced gluon spectrum off a high energy parton with energy $E$ \cite{Baier:1996sk,Baier:1996kr,Zakharov:1997uu,Arnold:2002ja,Wiedemann:2000za} (see also \cite{Mehtar-Tani:2013pia,Blaizot:2015lma} for recent reviews), 
\beq\label{eq:spectrum} 
\omega\frac{\rmd I}{\rmd\omega} &=&  \frac{\alpha_s C_R}{\omega^2} \, 2\rmR  \int_0^\infty \rmd t_2 \int_0^{t_2}\rmd t_1 \, \bdel_x \cdot \bdel_y\,  \Big[\Kc(\x,t_2|\y,t_1)- \Kc_0(\x,t_2|\y,t_1) \, \Big]_{\x=\y=\0}\,,\nn
\eeq
where the gluon is assumed to be soft, i.e. $\omega \ll E$ (cf.  Fig~.\ref{fig:gluon-radiation} ). We will relax this assumption in Section \ref{sec:generalization}.

The second term in \eqn{eq:spectrum} subtracts the vacuum part of the Green's function $\Kc$ which is solution of the Schr\"odinger equation:
\beq\label{eq:full-schordinger-0}
\left[ i \frac{\del}{\del t }  +\frac{\bdel^2}{2\omega} +i v(\x) \right] \Kc(\x,t|\y,t_1) =i \delta(\x-\y)\delta(t-t_1),
\eeq
where the imaginary potential $iv(\x)$ accounts for the random kicks that the energetic parton as well as the radiated gluon experience while traversing the plasma. It is related to the in-medium elastic cross-section by a Fourier transform,
\beq
v (\x,t) = N_c\int_\q\, \, \frac{\rmd\sigma_\text{el}}{\rmd^2 \q} \left(1-\rme^{i\q\cdot\x}\right),
 \eeq
where $\q$ is the transverse momentum exchange with the medium. The screening of the infrared (Coulomb) divergence depends on medium properties. In a thermal plasma, the HTL (Hard-Thermal-Loop) approximation yields \cite{Aurenche:2002pd} 
\beq \label{eq:el-HTL}
 \frac{\rmd^2 \sigma_\text{el}}{\rmd^2 \q} \equiv \frac{g^2 m^2_DT}{\q^2 (\q^2+m_D^2)},
\eeq
where $m^2_D= (1+N_f/6) g^2 T^2$ is the QCD Debye mass squared, with $N_f$ the number of active quark flavors and $T$ the plasma temperature.  In other approaches \cite{Gyulassy:2000er}, static scattering centers, introduced by Gyulassy and Wang (GW) \cite{Wang:1991xy}, are assumed. This corresponds to the following elastic cross-section,
\beq \label{eq:el-GW}
 \frac{\rmd^2 \sigma_\text{el}}{\rmd^2 \q} \equiv \frac{g^4 n}{(\q^2+\mu^2)^2},
\eeq
where $n$  is the density of scattering centers ($n\sim T^3$ for a medium in thermal equilibrium). 

The above two models for the elastic cross-section differ only when the moment transfer is of order the cut-off $\mu$.  The sensitivity of the gluon spectrum to the latter theoretical uncertainties is expected to be weak for dense enough  media where, due to multiple scatterings, the relevant transverse momentum scales in much larger: $1/x_\perp^{2}\sim Q^2 \sim \mu^2 N_\scatt \sim g^4 n L \gg \mu$. This reduced infrared sensitivity  is a consequence of the weak $\mu$-dependence of the large Coulomb logarithm:
\beq\label{eq:dipole-log}
v(\x,t)  \equiv \frac{1}{4} \hat q(\x^2,t)\, \x^2 \simeq \frac{g^4}{16\pi} N_c\,n(t) \, \x^2 \, \ln\frac{1}{\mu^2\,\x^2 }.
\eeq
Here, we have used the leading logarithmic definition of the transport coefficient \eqn{eq:qhat-def-lo}.  

For later convenience, we define the transport coefficient stripped of its Coulomb logarithm 
\beq\label{eq:qhat-stripped}
\qin(t)\equiv  4\pi \alpha_s^2 N_c \, n(t). 
\eeq 
From \eqn{eq:qhat-def-lo}, one can infer that $\qin= \hat q (Q^2=e \mu^2)$. 

In the absence of interactions the Green's function $\Kc$ reduces the free propagator
\beq 
 \Kc(\x,t_2|\y,t_1)= \Kc_0(\x-\y;t_2-t_1) =\frac{\omega}{2\pi i(t_2-t_1)}  \exp\left[i\frac{\omega(\x-\y)^2}{2(t_2-t_1)}\right],
\eeq
which reads
\beq\label{eq:FT-K0}
\int_{\q,\p}\,\Kc_0(\x-\y;t_2-t_1) \, \,\rme^{i\x\cdot\p-i\y\cdot \q}\,= (2\pi)^2\delta^{(2)}(\p-\q)\, \exp\left[-i\frac{\p^2}{2\omega} (t_2-t_1)\right],
\eeq
in momentum space.

Before we go on to discuss our method we shall first recall the main analytic limits, that are the basis of some phenomenological works, in somewhat more details than the above parametric discussion. 

\subsection{Single hard scattering approximation: Gyulassy-Levai-Vitev (GLV)} \label{sec:GLV}
In a dilute medium, the probability to undergo one scattering is very small and it is sufficient to expand \eqn{eq:spectrum} to leading order in $v(\x)$ ($N=1$ order in opacity), as follows \cite{Gyulassy:2000er,Wiedemann:2000za}
\beq\label{eq:n1-Kernel}
&&\bdel_x \cdot \bdel_y\, \Kc^{(N=1)}(\x,t_2|,\y,t_1)\Big|_{\x=\y=\0} = \nn
&& \bdel_x \cdot \bdel_y\,  \int \rmd \z \int_{t_1}^{t_2}\rmd s \, \Kc_0(\x-\z;t_2-s)  \,v(\z,s) , \Kc_0(\z-\y;s-t_1)\Big|_{\x=\y=\0}\nn
&=&  \int_{t_1}^{t_2}\rmd s \int_{\p,\q}   \,\p\cdot(\p-\q)\,\,\rme^{-i\frac{\p^2}{2\omega} (t_2-s)}  v(\q,s) \, \rme^{-i\frac{(\p-\q)^2}{2\omega} (s-t_1)},
\eeq
where we have performed a Fourier transform using \eqn{eq:FT-K0} and 
\beq
v(\q,s) \equiv  N_c\left[\frac{\rmd^2 \sigma_\text{el}}{\rmd^2 \q} - \delta(\q) \, \int_{\q'}\frac{\rmd^2 \sigma_\text{el}}{\rmd^2 \q'} \right].
\eeq
The integrations  over $t_1$ and $t_2$ are now straightforward:
\beq\label{eq:n1-time-integrals}
&&2 \rmR \int_0^\infty \rmd t_2 \int_0^{t_2} \rmd t_1 \int_{t_1}^{t_2}\rmd s \int_{\p,\q}   \,\p\cdot(\p-\q)\,\,\rme^{-i\frac{\p^2}{2\omega} (t_2-s)}  v(\q,s) \, \rme^{-i\frac{(\p-\q)^2}{2\omega} (s-t_1)} \nn
&& = 8 \omega^2\int_{0}^{\infty}\rmd s  \, \int_{\p,\q}  \, \,\frac{\p\cdot(\p-\q)}{\p^2(\p-\q)^2}\,\,   \,v(\q,s) \left\{1-\cos\left[\frac{(\p-\q)^2}{2\omega}s\right]\right\}.
\eeq
The term proportional to $\p^2$ in the last line of \eqn{eq:n1-time-integrals} vanishes owing to the fact that $\int_\q\, v(\q,s)=0$. On the other hand, $\delta(\q)$ in $v(\q,s)$ does not contribute to the term proportional to $\p\cdot \q$. It follows that
\beq
 \omega\frac{\rmd I_\GLV}{\rmd \omega}&=&  32 \pi \, \alpha_s \,C_R\,\qin\,  \int_{0}^{L}\rmd s  \, \int_{\p,\q}  \, \,\frac{\p\cdot\q}{\p^2(\p-\q)^2 (\q^2+\mu^2)^2}\, \left\{1-\cos\left[\frac{(\p-\q)^2}{2\omega}s\right]\right\},\nn
\eeq
where we have used the potential  (\ref{eq:el-GW}) and assumed that the color charges are uniformly distributed in a medium of length $L$, i.e., $n(t)=n \Theta(L-t)$. 

\begin{figure}[]
\center
\includegraphics[width=14cm]{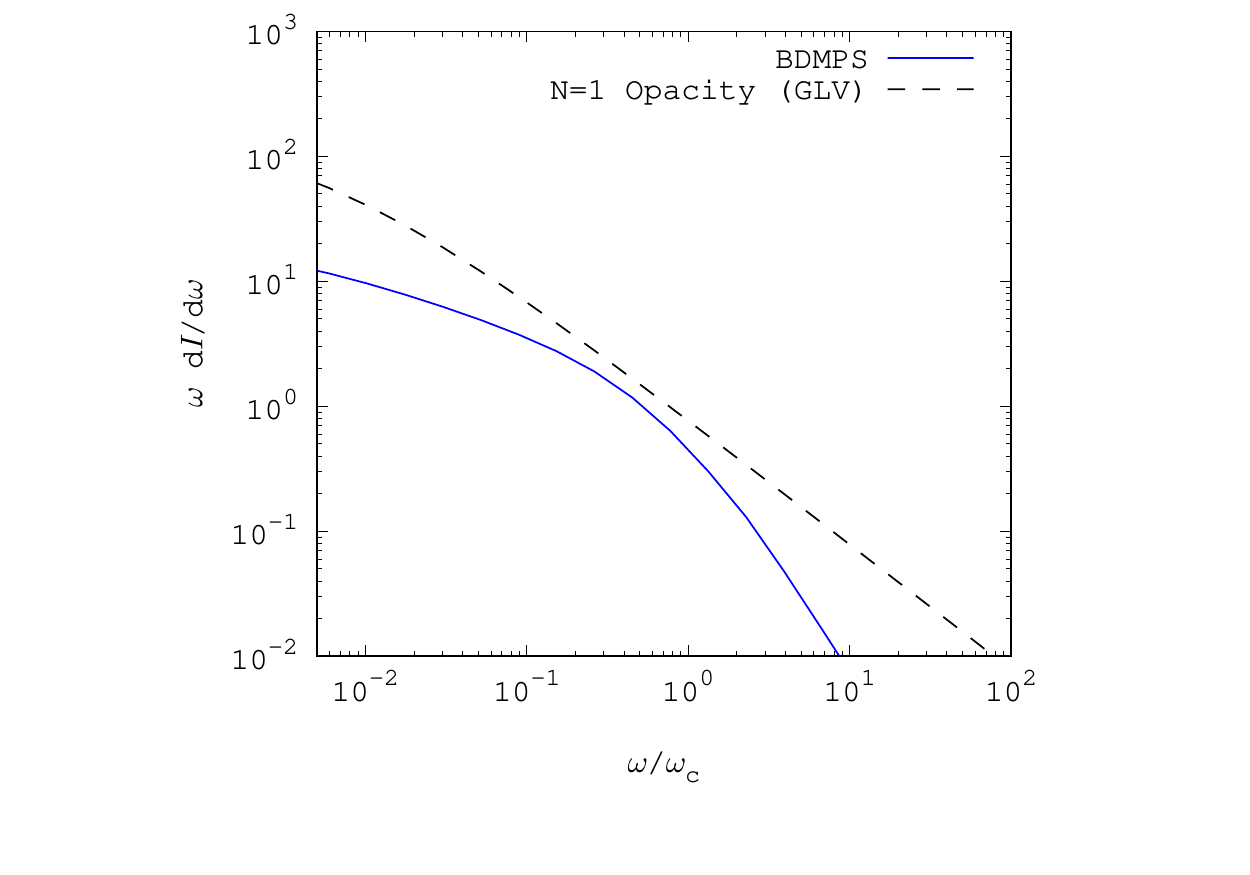}
\caption{The medium-induced spectrum to leading order in opacity (GLV) (dashed) and in multiple soft scattering approximation (BDMPS) (solid blue) computed for the following set of parameters: $\qin=0.1$ GeV$^{3}$, $L=4$ fm, $\mu=0.3$ GeV and we have set the overall pre-factor $\abar=1$. This corresponds to $\hat q \sim \qin  \ln (\qin L / \mu^2) = 1.5$ GeV$^2$/fm, $\omega_\BH \simeq 0.08$ GeV and $\omega_c \equiv \qin L^2 =40$ GeV. }\label{fig:spectrum-bdmps-glv}
\end{figure}

Making the following change of variables: $\k=\p-\q$, $\k \to \k \sqrt{2\omega/L}$ and $\p \to \p \sqrt{2\omega/L}$, one can simplify further 
\beq
&& \omega\frac{\rmd I_\GLV}{\rmd \omega}=  \frac{\abar\qin L^2 }{
\omega}  \int_{0}^{1}\rmd x  \, \int_0^{2\pi} \frac{\rmd \varphi}{2\pi} \int_0^\infty   \frac{\rmd k^2}{k^2} \nn
 && \qquad \qquad\times\int_0^\infty   \rmd p \, \,\frac{\del}{\del p}\frac{1}{(k^2+p^2-2 p k \cos \varphi+y)^2} \left[1-\cos\left(k^2  x \right)\right]. 
\eeq
Finally, the $p$, $\varphi$ and $x$ integrals can be carried out analytically yielding \cite{Salgado:2003gb}
\beq\label{eq:GLV}
 \omega\frac{\rmd I_\GLV}{\rmd \omega}&=& \frac{ \abar\qin\,L^2}{\omega}\,\int_0^\infty \rmd u\, \,\frac{1}{u+y} \,\frac{u-\sin(u)}{u^2}\,,
\eeq
where
\beq 
y\equiv\frac{\mu^2 L }{2 \omega}\,,
\eeq
and 
\beq
\abar \equiv\frac{ \alpha_s C_R}{\pi}\,.
\eeq
The GLV spectrum (\ref{eq:GLV}) is plotted in Fig.~\ref{fig:spectrum-bdmps-glv}. 
It encompasses two regimes characterized by the scale $\mu^2L/2$: 
\beq\label{eq:GLV-limits}
\omega\frac{\rmd I_\GLV}{\rmd \omega} \simeq  \abar \frac{\qin\,L}{\mu^2}
\begin{dcases}
\,\,\ln\left(\frac{\mu^2 L }{2 \omega} \right)\, \qquad\text{for} \quad \omega\, \ll\,\frac{1 }{2 } \,\mu^2 L\\
\,\,\frac{\pi}{4} \left(\frac{\mu^2 L }{2 \omega} \right) \quad\quad\text{for} \quad \omega\, \gg \,\frac{1 }{2 }\, \mu^2 L\\
\end{dcases}
\eeq

As we have alluded to in the parametric analysis performed in the previous section, another scale arises due to multiple scattering, $\omega_c \sim \qin L^2 \gg \mu^2 L $, and modifies the behavior of the spectrum in the infrared, for $\omega \ll \omega_c$.

\subsection{Multiple soft scattering approximation: Baier-Dokshitzer-Mueller-Peigne-Schiff  (BDMPS) }\label{sec:bdmps}
Let us now turn to the multiple soft scatting approximation often referred to as BDMPS (or Armesto-Salgado-Wiedemann (ASW) model \cite{Salgado:2003gb,Armesto:2003jh}). By neglecting the logarithmic dependence of the dipole cross section in \eqn{eq:dipole-log} assuming a constant scale $Q^2 \sim 1/\langle x_\perp^2\rangle $, the problem reduces to that of a harmonic oscillator (HO)  with a time dependent imaginary frequency $\Omega(t)$: 
\beq
iv(\x,t)\simeq \frac{i}{4} \bar q(t) \ln\frac{Q^2}{\mu^2} \x^2 \equiv  - \frac{\omega \Omega^2(t)}{2} \x^2,
\eeq
where 
\beq\label{eq:Omega-qhat}
\Omega^2(t) = - \frac{i}{2}\frac{ \qin (t) }{\omega}\, \ln\frac{Q^2}{\mu^2} \equiv - \frac{i}{2}\frac{\hat q (t)}{\omega}\,.
\eeq
That is
\beq\label{eq:im-omega}
\Omega (t)= \frac{1-i}{2} \sqrt{ \frac{\hat q (t) }{\omega}}.
\eeq
In terms of these new variables \eqn{eq:full-schordinger-0} reads: 
\beq\label{eq:ho-schordinger-2}
\left[ i \frac{\del}{\del t }  + \frac{\bdel^2}{2\omega} - \frac{\omega \Omega(t)^2}{2}\x^2\right] \Kc_\HO(\x,t|\y,t_0) =i \delta(\x-\y)\delta(t-t_0).
\eeq
Its solution reads \cite{Baier:1998yf,Abramowitz}
\beq
 \Kc_\HO(\x,t|\y,t_0) =\frac{i\omega}{2 \pi S(t,t_0)} \exp\left[ i S_\text{cl}(\x,t|\y,t_0)\right]\,,
\eeq
where the classical action and corresponding trajectory are defined as 
\beq \label{eq:cl-action}
S_\cl (\x,t|\y,t_0)=\frac{\omega}{2 }  \r_\cl [\xi] \frac{\rmd }{\rmd \xi} \r_\cl [\xi] \Big|_{t_0}^t\,,
\eeq
and 
\beq\label{eq:classical-path}
\left[\frac{\rmd^2}{\rmd\xi^2} +\Omega^2(\xi)\right] \r_\cl[\xi]=0,
\eeq
respectively. The determinant $S(t,t_0)$ obeys the same wave equation:
\beq\label{eq:wave-equation}
 \left[\frac{\rmd^2}{\rmd\xi^2} +\Omega^2(\xi)\right] S(\xi,t_0)=0,
\eeq
with boundary conditions $S(t_0,t_0)=0$ and $\del_\xi S(\xi,t_0)|_{\xi=t_0}=1$.

Given these boundary conditions the solution of \eqn{eq:classical-path} reads:
\beq\label{eq:classical-path-sol}
\r_\cl[\xi]= \frac{S(\xi,t_0)\,  \x +S(t,\xi) \, \y}{S(t,t_0)} 
\eeq
Inserting (\ref{eq:classical-path-sol}) into the classical action (\ref{eq:cl-action}) we find after some straightforward algebra,
\beq
 && S_\cl (\x,t|\y,t_0)= \frac{\omega}{2 S(t,t_0)} \nn
 && \times \left[ \del_t S(t,t_0) \, \x^2 - \del_{t_0} S(t,t_0)\, \y^2 +(\del_\xi S(t,\xi)|_{\xi=t} -\del_\xi S(\xi,t_0)|_{\xi=t_0})\,\x\cdot \y\right]\,.
\eeq
Adopting the notations introduced in \cite{Arnold:2008iy} we define the function 
\beq
C(t,\xi) = -  \del_{\xi } S(t,\xi) = \del_{\xi } S(\xi,t)\,,
\eeq
that corresponds to the other solution to \eqn{eq:wave-equation} with boundary condition $C(t,t)=1$. 
It follows that
\beq
 \del_\xi S(t,\xi)|_{\xi=t} = - C(t,\xi)|_{\xi=t} = - 1\,\and  \del_\xi S(\xi,t_0)|_{\xi=t_0} =   C(t_0,\xi)|_{\xi=t} =1\,,\nn
\eeq
where we have used the antisymmetric nature of $S(t,t_0)$, $S(t,t_0) = -S(t_0,t)$.
Finally, we obtain
\beq\label{eq:green-fct-sol}
 \Kc_\HO(\x,t|\y,t_1) =\frac{i\omega}{2 \pi S(t,t_0)} \exp\left[  i  \frac{\omega}{2 S(t,t_0)} \left[C(t_0,t) \, \x^2 + C(t,t_0)\, \y^2 -2 \, \,\x\cdot \y\right]\right]\,.\nn
\eeq
In the  time independent case, $\Omega(t) \equiv \Omega$, the $S$ and $C$ functions reduce to 
\beq\label{eq:sin-time}
S(t,t_0) = \frac{1}{\Omega} \sin\left(\Omega (t-t_0)\right)\quad \text{and } \quad C(t,t_0) = \cos\left(\Omega (t-t_0)\right)\,.
\eeq
By analyzing the argument of the exponential in \eqn{eq:green-fct-sol} using \eqn{eq:sin-time} one sees that the typical transverse scale in multiple-scattering regime is given by 
\beq\label{eq:x2-Omega}
\langle x_\perp^2\rangle \sim  \omega \Omega\sim  \frac{1}{\sqrt{\omega\hat q }}\,,
\eeq
which is the expected transverse momentum of medium-induced gluons $\langle k^2_\perp\rangle \sim  \langle x_\perp^2\rangle^{-1} \sim  \sqrt{\omega\hat q } $. This allows us to estimate the scale $Q^2$ using \eqn{eq:Omega-qhat} in \eqn{eq:x2-Omega}  (see detailed discussion in \cite{Arnold:2008zu})
\beq\label{eq:Q2}
Q^2 \sim \langle x_\perp^2\rangle^{-1} \simeq  \sqrt{\omega\hat q } = \sqrt{\omega \qin \ln(Q^2/\mu^2) }\,.
\eeq
Since for $Q^2 \gg \mu^2$ the logarithm is a slowly varying function $Q^2$ one can iterate the above equation  to obtain the leading logarithmic contribution: 
\beq \label{eq:Q2-estimate}
Q^2 \simeq  \sqrt{\omega \qin\ln(\sqrt{\omega \qin}/\mu^2) }\,.
\eeq
This estimate is valid in the multiple-scattering regime. For $\omega > \omega_c=\hat q   L^2/2$, the scale $Q^2$ is given by the maximum transverse momentum squared that multiple soft scatterings can transfer to the radiated gluon, namely, 
\beq\label{eq:Q2-estimate-2}
Q^2  \simeq Q^2_s \sim\qin L  \ln\left(\frac{\qin L}{\mu^2}\right)\,.
\eeq
Inserting \eqn{eq:green-fct-sol} in \eqn{eq:spectrum} yields
\beq\label{eq:spectrum-HO}
 \omega\frac{\rmd I_\HO}{\rmd \omega} = \frac{\alpha_s C_R}{\pi} 2\rmR  \int_0^\infty \rmd t_2 \int_0^{t_2}\rmd t_1 \,  \, \left[\frac{1}{S^2(t_2,t_1)} - \frac{1}{(t_2-t_1)^2}\right]\,.
\eeq
Using the following property \cite{Arnold:2008iy}
\beq\label{eq:dcot}
\del_t \left(\frac{C(t,s)}{S(t,s)}\right) =-\frac{1}{S^2(t,s)}\,,
\eeq
the $t_2$ integration can be performed and reads
\beq
 \int_{t_1}^\infty \rmd t_2 \, \frac{1}{S^2(t_2,t_1)} = \frac{C(t_1,t_1)}{S(t_1,t_1)}-\frac{C(\infty,t_1)}{S(\infty,t_1)}\,.
\eeq
The first term cancels against the vacuum piece, i.e., the second term in \eqn{eq:spectrum-HO},  while the second one can be integrated further over $t_1$ using \eqn{eq:S-C-composition}:
\beq
 \int_{0}^\infty \rmd t_1 \frac{C(\infty,t_1)}{S(\infty,t_1)} =-  \int_{0}^\infty \rmd t_1 \frac{\del_{t_1}C(t_1,L)}{C(t_1,L)}  = \ln C(0,L) = \ln \cos(\Omega L).
\eeq
Inserting the latter in \eqn{eq:spectrum-HO}  yields the BDMPS-Z result 
\beq\label{eq:bdmps-1}
  \omega\frac{\rmd I_\HO}{\rmd \omega} = 2 \abar  \ln |\cos(\Omega L)|\,.
\eeq
\eqn{eq:bdmps-1} encompasses two regimes
\beq
 \omega\frac{\rmd I_\HO}{\rmd \omega} &=& 2 \abar  \ln \left|\cos\left(\frac{1-i}{2}\sqrt{\frac{2 \omega_c}{\omega}}\right)\right|\simeq  2 \abar
 \begin{dcases}
\,\,\sqrt{\frac{\omega_c}{2\omega}}   \quad\qquad\text{for}\quad \omega \ll \omega_c\\
\,\,\frac{1}{12} \left(\frac{\omega_c}{\omega} \right)^2\quad\text{for} \quad \omega \gg \omega_c\\
\end{dcases}
\eeq
expressed in terms of the characteristic frequency 
\beq
\omega_c =\frac{1}{2}\hat q L^2\,.
\eeq
As discussed in the introduction of this section, the $\omega^{-1/2}$ is a consequence of the LPM effect for gluon radiation that occurs inside the medium corresponding to coherence time $t_\coh =\omega/k_\perp^2\ll L$. During the coherent scattering time $t_\coh$, the gluon acquires a transverse momentum squared of order $k_\perp^2 \sim \hat q \, t_\coh  $. 
At large formation times, and equivalently for $\omega >\omega_c$,  the spectrum is determined by a single hard scattering  characterized by the $k_\perp^{-4}$ power spectrum. In Sec.~\ref{sec:GLV} (cf. \eqn{eq:GLV-limits}), we have analyzed this regime and showed that it leads to the power spectrum $\omega^{-1}$. It follows that the multiple soft scattering contribution, $\omega^{-2}$, is negligible.  This so because, ignoring the Coulomb logarithm in MSSA, results in a Gaussian transverse momentum distribution that falls more rapidly than the actual power spectrum. Therefore, at high frequency the GLV-HT spectrum is the correct one.

In Fig.~\ref{fig:spectrum-bdmps-glv}, we have plotted the BDMPS spectrum (\ref{eq:bdmps-1}) in comparison with the GLV spectrum. The hard scale is calculated using \eqn{eq:Q2-estimate} for $\omega<\omega_c$ and \eqn{eq:Q2-estimate} for $\omega >\omega_c$ with the additional shift: $Q^2 \to Q^2 +\mu^2$ to enforce the condition $Q^2 >\mu^2$. The choice of parameters is such that the medium is dense: $Q_s^2= \hat q L \simeq 6 $ GeV$^2$, and the opacity parameter that enters the Coulomb logarithm is much larger than one, $Q_s^2/\mu^2 \simeq 67$. It is also motivated by phenomenological studies \cite{Burke:2013yra}. In this situation, the BDMPS spectrum leads to a larger LPM suppression than the GLV spectrum in the soft sector, $\omega < \omega_c$, its regime of validity. 

\section{Twist expansion in the Molière approximation} \label{sec:moliere}

\begin{figure}
\center
\includegraphics[width=0.4\textwidth]{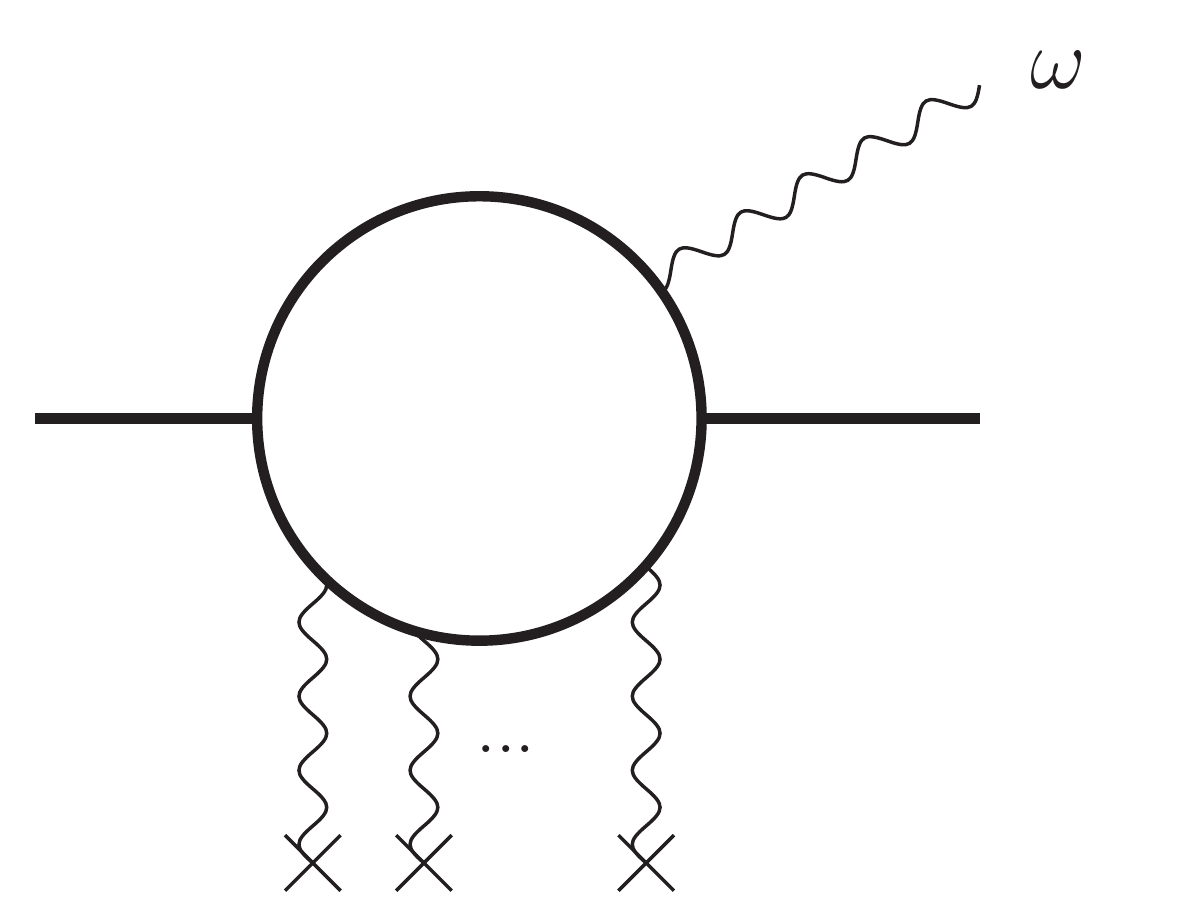}\qquad\includegraphics[width=0.4\textwidth]{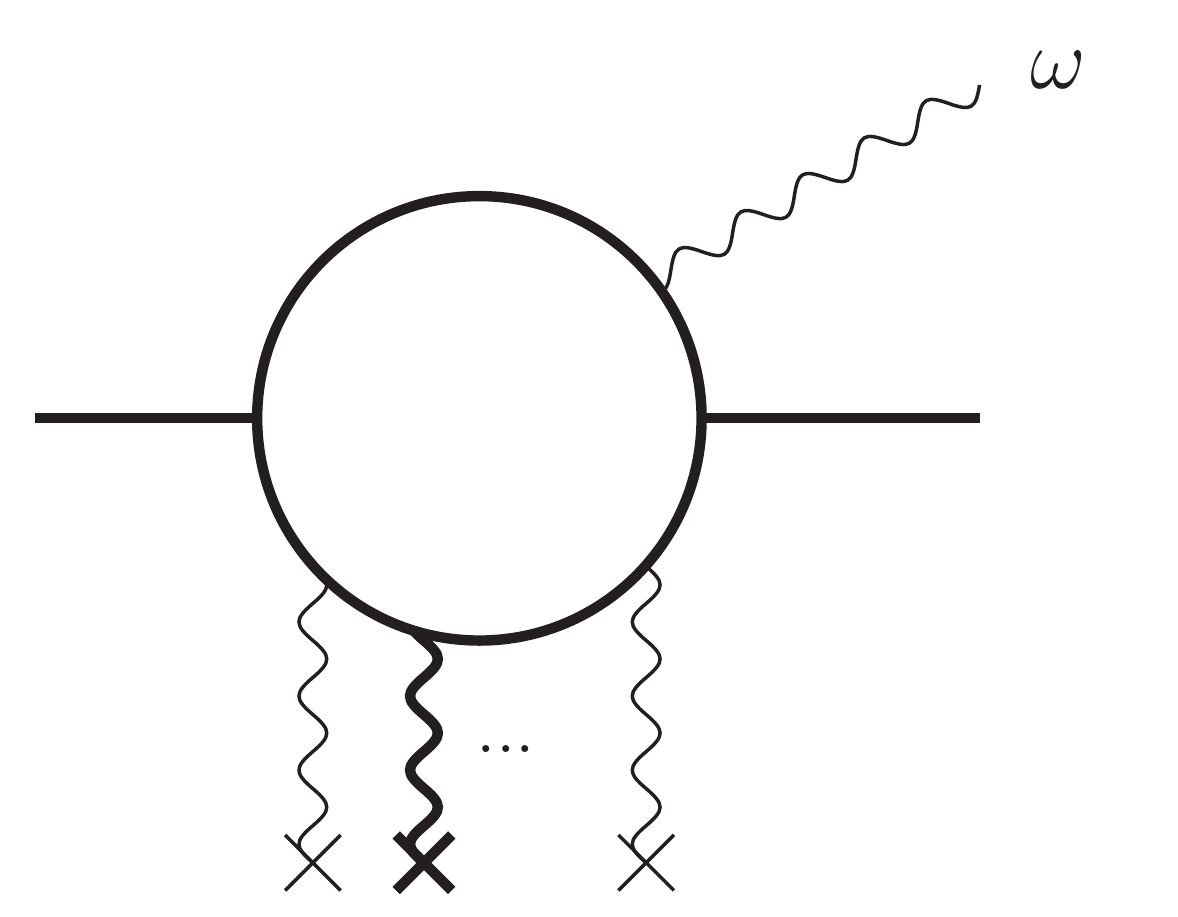}
\caption{A sketch of the first two orders in the modified opacity expansion. Left: the zeroth order where only multiple soft scatterings are accounted for (thin wavy vertical lines). Right: The NLO contribution, where one hard scattering is included (thick wavy line) while multiple soft scatterings are resummed to all orders. }
\label{fig:gluon-soft}
\end{figure}
In order to have a complete description of medium-induced gluon radiation our approach is to expand around the harmonic oscillator, such that the leading order reproduces the BDMPS approximation 
\beq
\omega \frac{\rmd I^{(0)}}{\rmd \omega} \equiv \omega \frac{\rmd I_\HO}{\rmd \omega} \,,
\eeq
and the next-to-leading order (NLO), $ \rmd I^{(1)}$, accounts for the single hard scattering limit as depicted in Fig.~\ref{fig:gluon-soft}.

To do so, we assume that $x_\perp \sim Q$, where the transverse scale  $Q$ is related to the typical transverse momentum broadening of the gluon during the radiation process (cf.  \eqn{eq:Q2}) and decompose the scattering potential as follows:
\beq
v(t,\x)=\frac{1}{4}\qin(t)\, \x^2 \left(\ln\frac{Q^2}{\mu^2} +\ln\frac{1}{\x^2 Q^2} \right)\equiv v_\HO(t,\x)+v_\pert(t,\x),
\eeq
and treat $v_\pert(t,\x)$ as a perturbation assuming 
\beq 
\ln\frac{Q^2}{\mu^2}  \, \gg\, \ln\frac{1}{\x^2 Q^2}. 
\eeq
\subsection{Next-to-leading order}
The Green's function $\Kc$ to leading order (LO) obeys the equation (see also \eqn{eq:ho-schordinger-2})
\beq\label{eq:ho-schordinger-3}
\left[ i \frac{\del}{\del t }  + \frac{\bdel^2}{2\omega} +i v_\HO(\x,t) \right] \Kc_\HO(\x,t|\y,t_1) =i \delta(\x-\y)\delta(t-t_1).
\eeq
The NLO obeys the inhomogeneous equation 
\beq\label{eq:NLO-schordinger}
\left[ i \frac{\del}{\del t }  + \frac{\bdel^2}{2\omega} +i v_\HO(\x) \right] \Kc_\pert(\x,t|\y,t_1) = - i v_\pert(\x)\,\Kc_\HO(\x,t|\y,t_1) .
\eeq
Using \eqn{eq:ho-schordinger-3} together with \eqn{eq:NLO-schordinger}, we can readily write 
\beq
 \Kc_\pert(\x,t_2|\y,t_1)  = - \int \rmd^2 \z \int_{t_1}^{t_2}\rmd s \, \Kc_\HO(\x,t|\z,s)  \,v_\pert(\z,s) \, \Kc_\HO(\z,s|\y,t_1). 
\eeq
Making use of the form (\ref{eq:green-fct-sol}) we can perform the transverse derivatives at the endpoints yielding 
\beq
\bdel_y \Kc_\HO(\z,s|\y,t_1) \Big|_{\y=\0}= -\frac{\omega^2}{2\pi\, S^2(s,t_1)} \, \z\, \exp\left[i \frac{ \omega }{2}\frac{C(t_1,s) }{S(s,t_1)} \, \z^2\right].
\eeq
To integrate over $t_1$ we shall use \eqn{eq:dcot}. Thus,
\beq
&&\int_0^s \rmd t_1\bdel_y \Kc_\HO(\z,s|\y,t_1) \Big|_{\y=\0}= - \int_0^s \rmd t_1 \frac{\omega^2}{2\pi\, S^2(s,t_1)} \, \z\, \exp\left[- i  \frac{\omega}{2 } \frac{C(t_1,s)}{S(t_1,s)} \, \z^2\right]\nn
&&\qquad= \frac{i \omega}{\pi\,\z^2 }\int_0^s \rmd t_1 \frac{-i\omega}{2\, }\,\z^2\,\del_{t_1} \left(\frac{C(t_1,s)}{S(t_1,s)}\right)  \, \z\, \exp\left[ -i  \frac{\omega}{2 } \frac{C(t_1,s)}{S(t_1,s)} \, \z^2\right]\nn
&&\qquad= \frac{i \omega}{\pi\, } \, \frac{ \z}{\z^2}\,  \left(   \exp\left[- i  \frac{\omega}{2} \frac{C(t_1,s)}{ S(t_1,s)} \, \z^2\right]\right)_{t_1=0}^{t_1=s}\nn
&&\qquad= - \frac{i \omega}{\pi\, } \,\frac{ \z}{\z^2}\,   \exp\left[- i  \frac{\omega}{2 } \frac{C(0,s)}{S(0,s)} \, \z^2\right],\nn
\eeq
where we have dropped the rapidly oscillating phase at $s=t_1$ in the last line. 
The other factor is evaluated in a similar fashion:
\beq
\bdel_x \Kc_\HO(\x,t_2|\z,s) \Big|_{\x=\0}= -\frac{\omega^2}{2\pi\, S^2(t_2,s)} \, \z\, \exp\left[ i  \frac{\omega}{2 } \frac{C(t_2,s)}{S(t_2,s)} \, \z^2\right]
\eeq
and 
\beq
&&\int_s^\infty \rmd t_2 \bdel_x \Kc_\HO(\x,t_2|\z,s) \Big|_{\y=\0}= - \int_s^\infty \rmd t_2 \frac{\omega^2}{2\pi\, S^2(t_2,s)} \, \z\, \exp\left[ i  \frac{\omega}{2 } \frac{C(t_2,s)}{S(t_2,s)} \, \z^2\right]\nn
&&\qquad= \frac{-i \omega}{\pi\, \z^2}\int_s^\infty \rmd t_2 \frac{i\omega}{2\, } \,\z^2\,\del_{t_2} \left(\frac{C(t_2,s)}{S(t_2,s)}\right)  \, \z\, \exp\left[i  \frac{\omega}{2 } \frac{C(t_2,s) }{S(t_2,s)}\, \z^2\right]\nn
&&\qquad= \frac{-i \omega}{\pi\, } \, \frac{ \z}{\z^2}\,  \left(   \exp\left[ i  \frac{\omega}{2 } \frac{C(t_2,s) }{S(t_2,s)}\, \z^2\right]\right)_{t_2=s}^{t_2=\infty}\nn
&&\qquad= \frac{-i \omega}{\pi\, } \, \frac{ \z}{\z^2}\,  \exp\left[ i  \frac{\omega}{2 } \frac{C(\infty,s)}{S(\infty,s)}\, \z^2\right]\,.\nn
\eeq
Putting the two pieces together, we obtain 
\beq\label{eq:spectrum-correction} 
\omega\frac{\rmd I^{(1)}}{\rmd\omega}&& = \frac{\alpha_s C_R}{\pi^2} \, 2 \rmR  \int_0^\infty \rmd s \,\int \,\frac{\rmd^2 \z}{\z^2} \, v_\pert(\z,s) \exp\left\{ -i \frac{\omega}{2 } \left[ \frac{C(0,s)}{S(0,s)} - \frac{C(\infty,s)}{S(\infty,s)}\right]\, \z^2\right\}\,. \nn 
\eeq
Now, we would like to rewrite $C(\infty,s)$ and $S(\infty,s)$ as a superposition of other solution to the wave equation \cite{Arnold:2008iy}:
\beq
S(t,t_1)=C(t_1,t_0) S(t,t_0)-S(t_1,t_0) C(t,t_0)\,,\nn
C(t,t_1)=-\del_{t_1}C(t_1,t_0) S(t,t_0)-\del_{t_1}S(t_1,t_0) C(t,t_0)\,.\nn
\eeq
Letting $t=\infty$, $t_1=s$ and $t_0=L$ yields
\beq\label{eq:S-C-composition}
S(\infty,s)=C(s,L) S(\infty,L)-S(s,L) C(\infty,L)\nn
C(\infty,s)=-\del_{s}C(s,L) S(\infty,L)+\del_{s}S(s,L) C(\infty,L)\nn
\eeq
In vacuum, $S(\infty,L) = \infty$. As a result only the first terms contribute to the ratio:
\beq
\frac{C(\infty,s)}{S(\infty,s)}=-\frac{\del_{s}C(s,L)}{C(s,L)} =  \Omega^2(s)\frac{S(s,L)}{C(s,L)}.
\eeq
Defining the ratio functions:
\beq
\Cot(t,t_0)= \frac{C(t,t_0)}{S(t,t_0)},\quad \Tan(t,t_0)= \frac{S(t,t_0)}{C(t,t_0)},
\eeq
we can further simplify the exponential in \eqn{eq:spectrum-correction}
\beq
 \exp\left\{ -i \frac{\omega}{2 } \left[ \frac{C(0,s)}{S(0,s)} - \frac{C(\infty,s)}{S(\infty,s)}\right]\, \z^2\right\}= \exp\left[ - i  \frac{\omega}{2 } (\Cot(0,s) - \Omega^2(s) \Tan(s,L)) \z^2\right].
\eeq
In the case of the brick, that is, $n(t)=n\Theta(L-t)$, we obtain
\beq
 \exp\left[  - i  \frac{\omega \Omega}{2} (\cot(\Omega s)-\tan(\Omega(L-s))) \z^2\right]
\eeq
In terms of the quantity 
\beq\label{eq:ks}
k^2(s)= i  \frac{\omega \Omega}{2} (\cot(\Omega s)-  \tan(\Omega(L-s))),
\eeq
\eqn{eq:spectrum-correction} reads
\beq\label{eq:spectrum-correction-2} 
\omega\frac{\rmd I^{(1)}}{\rmd\omega}&& = \frac{\alpha_s C_R}{\pi^2} \, 2 \rmR  \int_0^\infty \rmd s \,\int \,\frac{\rmd^2 \z}{\z^2} \, v_\pert(\z,s) \,\rme^{ - k^2(s) \z^2}\,. \nn 
\eeq
After making the following change of variables, $x = \z^2 Q^2  $, $\rmd^2 \z = (\pi/Q^2) \rmd x$, the integration over $x$ can be performed noting that
\beq 
\int_0^\infty \rmd x \, \ln\frac{1}{x} \,  \rme^{-a x}=\frac{1}{a} \left[ \ln(a) +\gamma\right],
\eeq
where $\gamma \approx 0.577$, is the Euler–Mascheroni constant.

As a final result, we find for the NLO contribution
\beq\label{eq:nlo}
\omega\frac{\rmd I^{(1)}}{\rmd\omega}&& = \frac{1}{2} \, \abar \, \qin \,\rmR  \int_0^L \rmd s \,  
\frac{1}{k^2(s)} \left[ \ln\frac{k^2(s)}{Q^2} +\gamma\right]\,.
\eeq
For completeness, let us show that we recover the GLV result at large frequency. When $\omega \to \infty$, we have  $\tan(\Omega(L-s))\to 0$ and $\cot(\Omega s)\to (\Omega s)^{-1}$. It follows that  $k^2(s) \approx i \omega /2 s$. Hence,
\beq
\frac{1}{2}\,\rmR  \int_0^L \rmd s \,  
\frac{1}{k^2(s)} \left[ \ln\frac{k^2(s)}{Q^2} +\gamma\right]  \,\rightarrow  \,\frac{\pi}{4} \frac{ L^2}{\omega},
\eeq
which yields the UV limit of \eqn{eq:GLV-limits}. 

In the IR, $\Omega \to (1-i)\infty$ as $\omega \to 0$,  and as result, $k^2(s) \to i\omega\Omega$. Therefore, \eqn{eq:nlo} reduces to 
\beq
 \omega\frac{\rmd I^{(1)}}{\rmd\omega}&& \simeq \frac{1}{2} \, \abar \, \qin \,\rmR  \int_0^L \rmd s \,\frac{2}{(1+i) \sqrt{\omega\hat q}} \left[ \ln\frac{(1+i) \sqrt{\omega \hat q }}{2 Q^2} +\gamma\right],\nn
 && \sim  \abar \, \sqrt{\frac{\hat q  L^2 }{ \omega }}\, \left(\frac{\qin }{\hat q }\right).\nn
  && \sim  \omega\frac{\rmd I^{(0)}}{\rmd\omega} \left(\frac{\qin }{\hat q }\right),
\eeq
where we have used that $Q^2 \sim \sqrt{\hat q \omega}$. 

\begin{figure}[]
\center
\includegraphics[width=14cm]{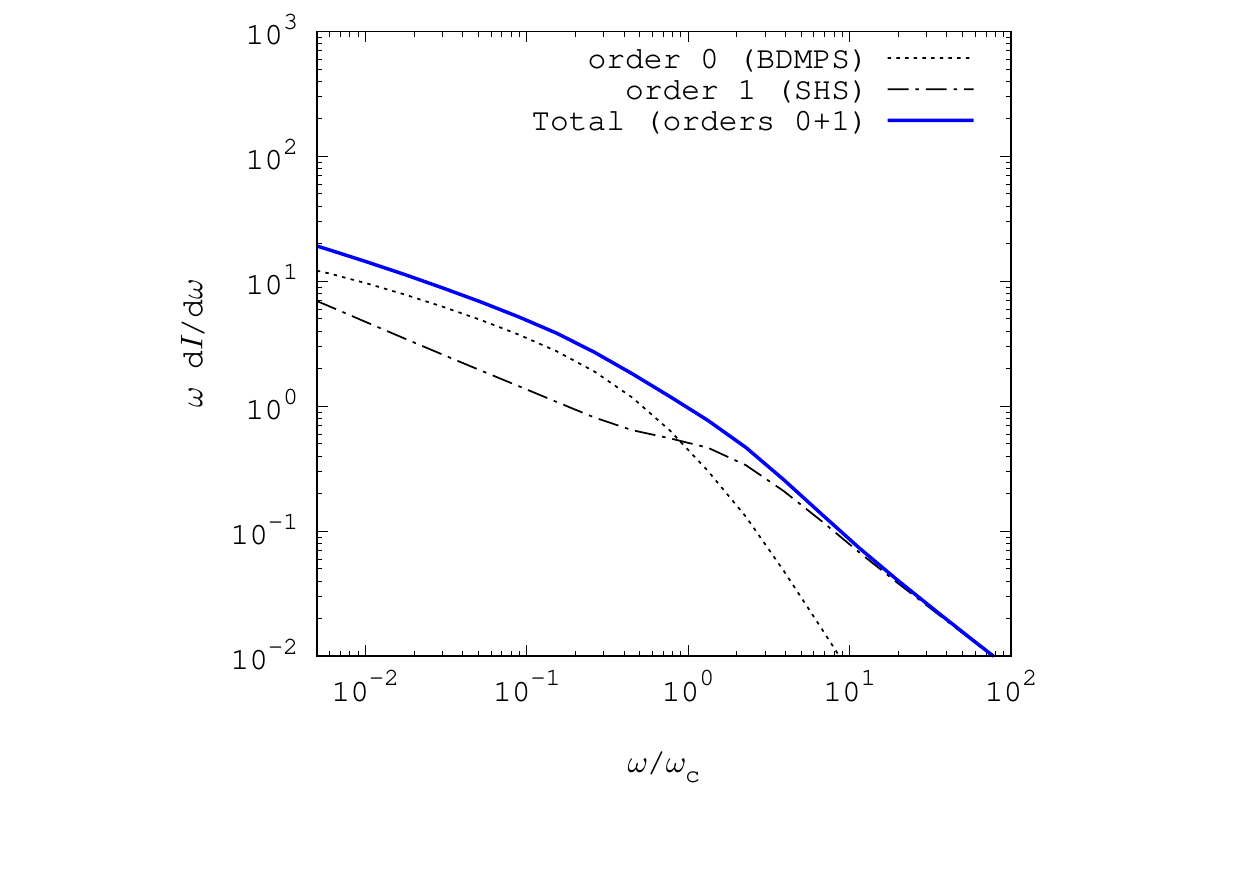}
\caption{The medium-induced spectrum to NLO in the expansion around the harmonic oscillator. The leading order (order 0) corresponds to the the BDMPS spectrum (dotted) and the first correction (order 1) accounts for to single hard scattering regime (dotted-dashed). The total spectrum is represented by a blue solid line. Numerical values of the parameters are the same as in Fig.~\ref{fig:spectrum-bdmps-glv}. }\label{fig:spectrum-new}
\end{figure}

\begin{figure}[]
\center
\includegraphics[width=14cm]{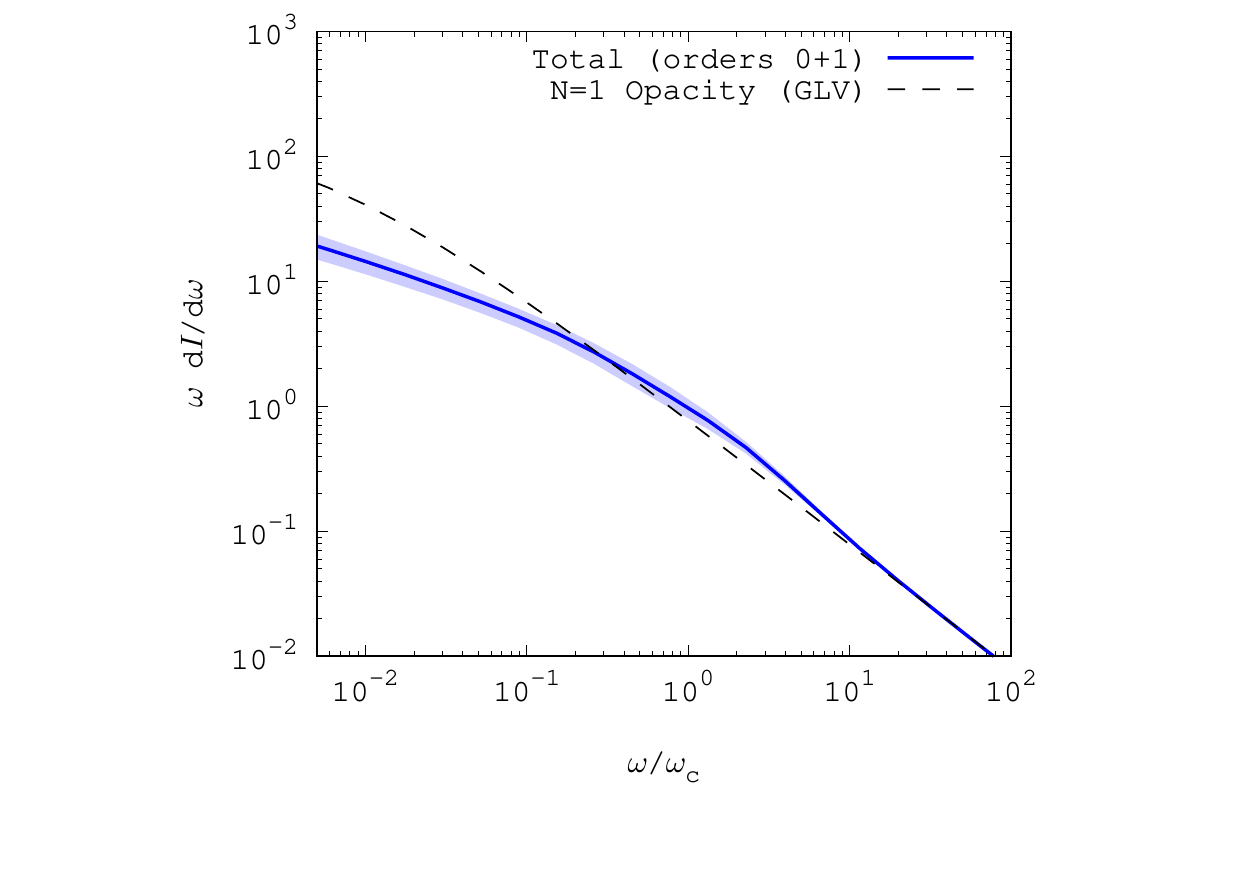}
\caption{The medium-induced spectrum to NLO in the expansion around the harmonic oscillator (BDMPS) (blue solid line). The blue band represents the variation of the IR transverse scale between $\mu^2/2$ and $2 \mu^2$, $\mu=0.3$ GeV being the central value. The GLV spectrum (dashed) is also plotted for comparison.}\label{fig:spectrum-total-glv}
\end{figure}

This implies that in the soft regime the NLO exhibits the same scaling as the LO, i.e., $\omega^{-1/2}$ , but is suppressed by one power of the Coulomb logarithm: $\qin / \hat q  \sim \ln^{-1}(\sqrt{\omega \hat q}/\mu^2)$ (cf. Figs.~\ref{fig:spectrum-new} and \ref{fig:spectrum-total-glv}).  

\subsection{Numerical results}
To summarize our results, the total spectrum reads 
\beq\label{eq:total-spectrum}
 \omega\frac{\rmd I}{\rmd \omega }\simeq 2\,\abar \, \ln |\cos(\Omega L)|  + \frac{1}{2}\,\abar\,\qin \, \rmR  \int_0^L \rmd s \,  
\frac{1}{k^2(s)} \left[ \ln\frac{k^2(s)}{Q^2} +\gamma\right].
\eeq
where 
\beq
Q^2 \simeq \sqrt{ \omega \hat q (Q^2)} \equiv  \sqrt{ \omega \qin \ln(Q^2/\mu^2)}\simeq \sqrt{ \omega \qin\ln(\sqrt{\omega \qin }/\mu^2)}  \
\eeq
and  $\Omega$ is given by \eqn{eq:im-omega}. 

\eqn{eq:total-spectrum} is our  main result. It is plotted in Fig.~\ref{fig:spectrum-new}, where we can see that the BDMPS approximation (order 0) dominates below $\omega_c\sim\hat q L^2$ and the spectrum exhibits the characteristic $\omega^{-1/2}$ scaling, while in the UV the single hard scattering is the leading contribution. In Fig.~\ref{fig:spectrum-total-glv} we compare our full result to the GLV spectrum and observe that they agree for $\omega \gg \omega_c$. We have also varied the IR transverse cut-off $\mu^2$, that appears in the Coulomb logarithm, by factors of $2$ and $1/2$  to gauge the sensitivity of the spectrum for  $\omega \gg \omega_\BH$ to modeling of the elastic cross-section $\rmd \sigma_\el(q_\perp)$ when $q_\perp\sim \mu$ as illustrated in \eqn{eq:el-HTL} and \eqn{eq:el-GW}.  We obtain about $\pm 20$\% variation in the multiple soft scattering regime, while in the UV the difference is not significant. The latter is to be expected since the transverse momentum is cut-off from below by the large transverse scale $\sqrt{\omega/L}$.

\section{Generalization to finite gluon energy }\label{sec:generalization}

So far, we have only addressed the case of soft gluon radiation for simplicity. Our approach, however, holds for arbitrary gluon frequency. 
As function of the energy fraction $z\equiv \omega /E$, the generalization beyond the soft gluon reads
\beq\label{eq:spectrum-z} 
z\frac{\rmd I}{\rmd z} &=&  \frac{\alpha_s\,z  P(z)}{ \left[z(1-z) E\right]^2}\,  \, \rmR  \int_0^\infty \rmd t_2 \int_0^{t_2}\rmd t_1 \, \bdel_x \cdot \bdel_y\,  \Big[\Kc(\x,t_2|\y,t_1)- \Kc_0(\x,t_2|\y,t_1) \, \Big]_{\x=\y=\0}\nn
\eeq
where $P(z)\equiv P_{gR}(z)$ is the Altarelli-Parisi splitting function for a parton in representation $R$ radiating a gluon with energy fraction $z$. Here, the Green's function $\Kc$ solves the following Schr\"odinger equation:
\beq\label{eq:full-schordinger}
\left[ i \frac{\del}{\del t } +\frac{\bdel^2}{2z(1-z)E} +iv(\x,z)\right] \Kc(\x,t|\y,t_1) =i \delta(\x-\y)\delta(t-t_1)\,,
\eeq
where the scattering potential reads
\beq
v(\x,z) \equiv \frac{1}{2} N_c \sigma(\x)+\left(C_R-\frac{1}{2} N_c \right) \sigma(z\x) + \frac{1}{2}  \sigma((1-z)\x)\,,
\eeq
with
\beq
\sigma(\x)\equiv  \int_\q \, \frac{\rmd\sigma_\text{el}}{\rmd^2 \q} \left(1-\rme^{i\q\cdot\x}\right)\,.
\eeq
Introducing once again the separation scale $Q^2$ that we shall estimate shortly, we can split the above potential into a leading logarithmic contribution to be resummed to all orders and a perturbative piece
\beq
v(\x,z)  = v_\HO(\x,z) + v_\pert(\x,z)\,.
\eeq
where
\beq 
v_\HO(\x,z)\simeq \frac{1}{8}\qin\, \x^2  \left[ 1 +\left(\frac{2C_R}{N_c }-1 \right)z^2 +  (1-z)^2 \right] \ln\frac{Q^2}{\mu^2} \equiv \frac{1}{4}\hat q_\eff\, \x^2 ,\nn
\eeq
We have introduced the effective transport coefficient
\beq
\hat q_\eff (Q^2) \equiv \frac{1}{2}\qin\,  \left[ 1 +\left(\frac{2C_R}{N_c }-1 \right)z^2 +   (1-z)^2 \right]\,\ln\frac{Q^2}{\mu^2}\,,
\eeq
where we have generalized to arbitrary flavors adopting the notations leading to Eqs.~(5.5) and (5.6) in \cite{Blaizot:2012fh}.
The typical branching time reads
\beq
t_\br \equiv  \sqrt{\frac{z(1-z) E}{\hat q _\eff}}.
\eeq
This allows to estimate the transverse scale $Q^2$ as follows 
\beq
Q^2 \equiv   \hat q_\eff\, t_\br= \sqrt{\frac{1}{2}z (1-z) E \qin\left[ 1 +\left(\frac{2C_R}{N_c }-1 \right)z^2 +   (1-z)^2 \right] \ln \frac{Q^2}{\mu^2}}.\nn
\eeq
Note that when $z =\omega/E \to0$, we recover the soft limit $Q^2 \sim \sqrt{\omega \qin \ln (Q^2 /\mu^2)}$. The perturbation part of the scattering potential reads
and 
\beq 
\label{eq:potential-pert-z}
&& v_\pert(\x,z)\simeq \frac{1}{8}\qin\x^2  \nn
&& \times\left[  \ln\frac{1}{\x^2Q^2}+\left(\frac{2 C_R}{N_c}-1 \right)z^2 \ln\frac{1}{z^2\x^2Q^2}+(1-z)^2 \ln\frac{1}{(1-z)^2\x^2Q^2}\right],\nn
\eeq
Now, using the results obtained for the soft limit it is straightforward to generalize to finite $z$. Modulo the Altarelli-Parisi splitting, the leading order is obtained by making the substitution  $\omega \to z(1-z)E$ and  $\hat q \to \hat q_\eff$,
\beq
z\frac{\rmd I^{(0)}}{\rmd z}  = \frac{\alpha_s}{ \pi }\,z P(z)\, \ln\left| \cos \Omega L\right|\,
\eeq
where 
\beq
 \Omega =\frac{i-1}{2} 
 \sqrt{\frac{\hat q_\eff(z,Q^2)}{z(1-z) E}}.
\eeq
The NLO is composed of three terms corresponding to the three terms in  \eqn{eq:potential-pert-z}. By introducing the function 
\beq
F(x) \equiv  \frac{1}{x} \left[ \ln(x) +\gamma\right]\,,
\eeq
we can write the correction to the HO as follows
\beq
z\frac{\rmd I^{(1)}}{\rmd z}  = \frac{\alpha_s}{ 4\pi }\, zP(z)\, \, \frac{\qin }{Q^2}  \, \rmR  \int_0^L \rmd s \,  \left\{F\left[\frac{k(s)}{Q^2}\right]+F\left[\frac{k(s)}{z^2Q^2}\right]+F\left[\frac{k(s)}{(1-z)^2Q^2}\right]\right\}.
\eeq
with again 
\beq\label{eq:ks-z}
k^2(s)= i  \frac{\omega \Omega}{2} (\cot(\Omega s)-  \tan(\Omega(L-s))).
\eeq

\section{Summary and outlook} \label{sec:summary}
Medium-induced gluon radiation in a dense QCD medium is at the core of jet-quenching physics. A part from numerical solutions, the spectrum is analytically known only in limiting cases. At low frequency, the scattering potential can be approximated by that of a harmonic oscillator by neglecting the variation of the Coulomb logarithm. This approximation, where multiple scatterings can be resummed to all orders, leads to a Gaussian distribution for transverse coordinates and momenta. As a result, the physics of hard scattering, characterized by the power spectrum $k_\perp^{-4}$,
 is not correctly accounted for and the approximation fails for $\omega \gg \omega_c $. 

In this work, we have developed a new approach to analytically account for both limits. It consists in expanding the  radiation spectrum around the harmonic oscillator (BDMPS approximation) and we have shown that the NLO encompasses the UV limit of leading order in opacity expansion (GLV approximation).  We have in particular demonstrated that this method is suited for analytic calculations and allows for systematically compute higher orders. 

We have focuses our discussion on the energy spectrum integrated over transverse momentum. In a future work, we plan to extend this approach by including transverse momentum dependence. Another improvement is account for the Bethe-Heitler regime by going beyond the leading-logarithmic approximation and take into account power corrections.

\section*{Acknowledgements} 
We would like to thank K. Tywoniuk for fruitful discussions and careful reading of the manuscript. This work is supported by the U.S. Department of Energy, Office of Science, Office of Nuclear Physics, under contract No. DE- SC0012704,
and in part by Laboratory Directed Research and Development (LDRD) funds from Brookhaven Science Associates.

,
\bibliographystyle{elsarticle-num}



\end{document}